\input harvmac.tex
\noblackbox

\Title{USC-99-002}
{\vbox{
\centerline{The continuum limit of $sl(N/K)$ }
 \vskip 4pt
\centerline{integrable super spin chains.}}}

\centerline{H. Saleur}

\bigskip\centerline{Department of Physics}
\centerline{University of Southern California}
\centerline{Los Angeles, CA 90089-0484}

\vskip .3in

I discuss in this paper the continuum limit of integrable spin chains
based on the superalgebras $sl(N/K)$. The general conclusion is that, with the 
full ``supersymmetry'', none of these models is relativistic. When the supersymmetry
is broken by the generator of the sub $u(1)$, Gross Neveu models of various types 
are obtained. For instance, in the case of $sl(N/K)$ with a typical fermionic representation on 
every site, the continuum limit is the GN model 
with $N$ colors
and $K$ flavors. In the case of $sl(N/1)$ and atypical representations of spin $j$, a close cousin of  the GN model
with $N$ colors and $j$ flavors with flavor anisotropy is obtained. The Dynkin parameter associated with the fermionic root, while providing solutions of 
the Yang Baxter equation with a continuous parameter, thus does not give rise to 
any new physics in the field theory limit.

These features  generalize to the case where
 an impurity is embedded in the system.

\Date{05/99}

\newsec{Introduction}

The continuum  (field theory) limit of 
quantum spin chains with ordinary symmetries, whether integrable or not, is generally well understood, and described by 
field theories whose  symmetries closely match the ones of the underlying lattice model. The most striking 
example of this phenomenon is furnished by the case of $sl(N)$ integrable spin chains, whose continuum 
limits are   $SU(N)$ Wess Zumino models, with a  level that depends on the representation used to 
build up the chain \ref\A{I. Affleck, Nucl. Phys. B265 (1986) 409; H. M. Babujian, Nucl. Phys. B215 (1983) 317.}. Quantum group deformations of the lattice symmetries are also known 
to give rise smoothly to similar deformations in the field theory and the associated 
scattering matrices. 

In contrast, the continuum limit of spin chains based on superalgebras is rather poorly understood. This is unfortunate,
since the question is related to physical problems of the highest interest, in particular in the 
context of disordered systems \ref\HR{F. D. M. Haldane and E. H. Rezayi, Phys. Rev. Lett. 60 (1988) 956.},
\ref\MR{M. Milovanovic and N. Read, Phys. Rev. B53 (1996) 13559.}, \ref\Z{M. R. Zirnbauer, J. Math. Phys. 38 (1997) 2007.},
and maybe of $N=2$ supersymmetry \ref\Z{Z. Maassarani, J. Phys. A28 (1995) 1305.}. One thing that seems clear, is that,
in the integrable case, none of these continuum limits have to do with  the corresponding Wess Zumino models on supergroups: this
is only expected, since these WZW models can present very pathological non unitary properties \ref\RoS{L. Rozansky and H. Saleur,
Nucl. Phys. B376 (1992) 461.},\ref\CKT{J.S. Caux, I. I. Kogan and A. M. Tsvelik, Nucl. Phys. B489 (1996) 444.}: if not,
what, then, are these continuum limits? The same question arises after quantum group deformation. Here, some
preliminary results have indicated a very rich structure: it was indeed shown in \ref\S{H. Saleur, hep-th/9811023, to appear in J. Phys. A Lett. } that the continuum 
limit of a particular $osp(2/2)_q$ model \ref\RMI{M. J. Martins, P. B. Ramos, Phys. Rev. B56 (1997) 6376.} coincided with the continuum limit 
of the well known Bukhvostov Lipatov model \ref\BL{A. P. Bukhvostov, L. N. Lipatov, Nucl. Phys. B180 (1981) 116.}, providing the first example of an integrable ``double sine-Gordon'' model
\ref\F{V. A.  Fateev, Nucl. Phys. B473 (1996) 509.}, \ref\LSS{F. Lesage, H. Saleur, P. Simonetti, 
Phys. Rev. B57 (1998) 4694.}.  The case of $sl(N/K)_q$ models 
based on fundamental representations has also been studied in some details \ref\V{H. de Vega and E. Lopes, Phys. Rev. Lett. 67 (1991) 489.}, enough to show that they have 
a relativistic limit, but far from providing a complete identification of the latter. 
It is thus pretty clear  that a bunch of solvable field theories
of the highest interest are lurking behind super spin chains, and this paper is a first step at clarifying the situation.

Putting the question of field theory aside, integrable lattice models based on super algebras
 have a rather long history, starting with 
the t-J model, which corresponds to $sl(2/1)$ and the fundamental representation. Generalized 
t-J models, based on $sl(N/1)$ and still the fundamental representation have also been studied
in the context of strongly interacting electrons, and - although maybe the algebraic 
origin of the models was not so clear - in the study of quantum impurity problems 
with the degenerate Anderson model. Following 
developments in superconductivity, models based on more complex algebras or representations
have been considered: for instance,
the model based on the algebra $sl(2/1)$ and typical four dimensional representations
was introduced in \ref\BGLZ{A. J. Bracken, M. D. Gould, J. R. Links and Y.Z. Zhang, Phys. Rev. Lett. 74 (1995) 2768;
Z. Maassarani, J. Phys. A. 28 (1995) 1305.}, while the
model based on $sl(2/2)$ and the fundamental (which is also typical) was introduced in \ref\EKS{F. Essler, V. Korepin and
K. Schoutens, Phys. Rev. Lett. 68 (1992) 2960.}.

The paper is organized as follows. In sections 2, 3 and 4, I introduce the 
general Bethe ansatz for chains based on typical fermionic representations. The ground  state and physical excitations are studied in section 6, while in section 7, I determine the S matrix and the mapping onto the Gross-Neveu model. Sections 7 and 8 are devoted to special cases, in particular 
 atypical fermionic representations and ``bosonic'' representations. The whole study is extended  in section 9 to the case of chains with impurities. A few conclusions are gathered in section 10. 

Before starting, I would like to stress that, although the continuum limit of these models has not been systematically studied before,
the following  has a some overlap with known partial results from \ref\TI{A. M. Tsvelik, Sov. Phys. JETP 66 (1987) 754.} and  \ref\FPT{H. Frahm, M. Pfannm\"uller, A. M. Tsvelik, Phys. Rev. Lett. 81 (1998) 2116; H. Frahm, cond-mat/9904157}. 

\newsec{The Bethe equations}

Recall that $sl(N/K)$ has $N+K-1$ Cartan generators, the first $N-1$ belonging
to $sl(N)$, the last $K-1$ to $sl(K)$, the special generator $H_N$ being associated with 
the odd root. The Dynkin diagram
decomposes into $sl(N)$ and $sl(K)$ parts, connected by the odd root:

\bigskip
\noindent
\vskip.2cm\hskip-3cm\centerline{\hbox{
\rlap{\raise28pt\hbox{\hskip.38cm
$\bigcirc$------$\bigcirc$- - - $\bigotimes$- - - $\bigcirc$------$\bigcirc$}}
\rlap{\raise39pt\hbox{$\hskip .3cm a_1\hskip.6cm a_2\hskip.7cm a_N\hskip1.8cm
a_{N+K-1}$}}}
}
\smallskip

\noindent I wish to  consider first  integrable hamiltonians with a {\sl fermionic}
representation of $sl(N/K)$ on every site. This is a priori the most interesting case,
since these representations exhibit, in the typical case, a continuous parameter -  a feature
that is  absent in models based on ordinary algebras. One might hope 
that this parameter 
describes some  interesting new physics - maybe giving rise to a multiparameter 
integrable quantum field theory. As we will see shortly, the detailed study of the 
exact solution does not support that expectation unfortunately.

The   Dynkin parameters of what I call (a bit incorrectly) fermionic 
representations are  $(0,\ldots,0,t,0,\ldots,0)$, $t$ being a real number, which I assume 
positive  in what follows: the case $t<0$ would
follow simply by exchanging $N$ and $K$. For $t$ generic, this representation is 
typical, with dimension $2^{NK}$, and vanishing super dimension. Atypical cases correspond to
 $t$ integer, $-(K-1)\leq t\leq N-1$.

I call the  roots of the Bethe equations
 $\mu_{N-1},\ldots,\mu_1,\mu_0,\lambda_1,\ldots,\lambda_{K-1}$, 
and introduce the function 
\eqn\keri{e_t(\nu)={\nu+it/2\over\nu-it/2}.}
The Bethe equations read then
\eqn\bethe{\eqalign{1=&\prod e_2(\mu_{N-1}-\mu'_{N-1})
e_{-1}(\mu_{N-1}-\mu_{N-2})\cr
1=&\prod e_{-1}(\mu_p-\mu_{p-1})e_{2}(\mu_p-\mu'_p)e_{-1}
(\mu_p-\mu_{p+1}),\ p=N-2,\ldots,1\cr
e_{t}^L(\mu_0)=&\prod e_{1}
(\mu_0-\mu_1)e_{-1}(\mu_0-\lambda_1)\cr
1=&\prod e_{-1}(\lambda_1-\lambda_2) e_2(\lambda_1-\lambda'_1)e_{-1}
(\lambda_1-\mu_0)\cr
1=&\prod e_{-1}(\lambda_p-\lambda_{p-1})
e_2(\lambda_p-\lambda'_p)e_{-1}(\lambda_p-\lambda_{p+1}),\ p=2,\ldots,K-2\cr
1=&\prod e_{-1}(\lambda_{K-1}-\lambda_{K-2})
e_{2}(\lambda_{K-1}-\lambda'_{K-1}),\cr}}
where as usual \ref\K{N. Yu Reshetikhin, Lett. Math. Phys. 14 (1987) 235.}, the pattern of $e$ labels reproduces the Cartan matrix: the two 
salient features are the absence of $\mu_0,\mu_0$ coupling, and the opposite couplings of $\mu_0$ to
$\mu_1$ and $\lambda_1$ respectively.  
The notation is obvious but implicit. For instance in the first 
equation the product is taken over all Bethe roots $\mu_{N-2}$ and all Bethe roots $\mu'_{N-1}$
different from $\mu_{N-1}$. 

To get some intuition about these equations, one can think of the dimension of the typical representation $2^{NK}$ as the number of possible  ways of 
putting 
fermions with $N$ colors and $K$ flavors on a given site of the chain. For instance in the case of $sl(2/1)$, the four states can be interpreted as empty, 
one fermion with spin up or down, and finally a pair of fermions. 
The parameter $N_{\mu_0}$ can be interpreted
as the number of fermions; the numbers of fermions with a given color
are then given by  
 $N_{\mu_0}-N_{\mu_1},N_{\mu_1}-N_{\mu_2},\ldots, N_{\mu_{N-2}}-N_{\mu_{N-1}}$,
$N_{\mu_{N-1}}$,
and the numbers of fermions with a given flavor by   
$N_{\mu_0}-N_{\lambda_1},N_{\lambda_1}-N_{\lambda_2},\ldots, N_{\lambda_{K-2}}-N_{\lambda_{K-1}}$, $N_{\lambda_{K-1}}$. Dynkin parameters can easily be deduced from this and the knowledge of the Cartan generators in the typical representation.

The energy takes the form
\eqn\bareenergy{E=\epsilon\sum_{\mu_0} {t\over \mu_0^2+{t^2\over 4}}+ A\sum_{\mu_0} 1,}
where
I have put a chemical potential for the number of fermionic Bethe roots, and $\epsilon=\pm 1$. 
Explicit expressions for the hamiltonians themselves  can be found in the 
references below for some special cases; they are, of course, quite intricate, except for the simplest values of $N$ and $K$.

I am not aware of a general derivation of equations \bethe, even though it is presumably possible using the 
general techniques developed in \ref\RM{P.B. Ramos and M. J. Martins, Nucl. Phys. B500 (1997) 579.}, and the form is very natural from algebraic considerations \K. 
A number of particular cases have already been studied however; besides $sl(2/1)$  \ref\PF{M. P. Ffannm\"uller, H. Frahm, Nucl. Phys. B479 (1996) 575.}
and $sl(2/2)$, recall that the 
fermionic representations
we are considering  can become atypical for special values of the parameter $t$. In the case
of $sl(N/1)$, the value $t=1$ corresponds in fact to the fundamental representation, and the model we are interested in
coincides then with the $su(N)$ t-J model which was extensively studied by Schlottmann \ref\LS{K. Lee and P. Schlottmann,
J. Physique Coll. 49 (1988) C8 709; P. Schlottmann, J. Phys. C4 (1992) 7565.}.  There also  exists by now a 
huge literature of quantum deformation of super groups and various considerations 
about graded inverse scattering method, with motivations ranging from   properties of electronic materials
to knot theory.

The solutions of the Bethe  equations are as follows. Consider first the
fermionic Bethe roots $\mu_0$. Because  there is no $(\mu_0,\mu_0)$ 
coupling on the right hand side of the Bethe equations (the corresponding
element of the Cartan matrix vanishes), the usual string 
solutions, well known for ordinary algebras,  are not possible. 
However because the coupling 
between $\mu_0$ and $\mu_1$
 has a  sign opposite to the one for ordinary algebras, it is possible
 to compensate for the growth or
 decay of the
left hand side of the Bethe equations when $\mu_0$ has an imaginary part
 by having
complexes of ``strings over strings''. Such complexes were probably first introduced 
by Takahashi \ref\T{M. Takahashi, Prog. Th. Phys. 44 (1970) 899.} in his study of 
one dimensional fermions interacting with an attractive delta 
function potential; they have been widely used since, in particular by Schlottmann in his 
study of models based on the fundamental of  $sl(N/1)$. They are of the type
\eqn\cplxs{\eqalign{\mu_0&=\mu^{p-1}+(-(p-1)i/2,\ldots,(p-1)i/2)\cr
\mu_q&=\mu^{p-1}+(-(p-1-q)i/2,\ldots,(p-1-q)i/2),\ q=1,\ldots, p-2\cr
\mu_{p-1}&=\mu^{p-1},\cr}}
for $p=1,\ldots,N$. Hence,   for $\mu_0$ there are string solutions of
 length smaller or equal to $N$. 
 Of course the patterns \cplxs\ are obeyed only in
the large $L$ limit. For finite $L$ the solutions of the Bethe equations differ
from \cplxs\  by exponentially small amounts. One has to use and eliminate these
small deviations to rewrite Bethe equations involving the complexes. As for the $\mu_p,\ p>0$
roots that
are not involved in such complexes and the $\lambda_p$ roots, 
they are determined by the same arguments
as for $sl(N)$ and $sl(K)$ respectively ie they can be strings of
 any possible length. Observe that the role of $sl(N)$ and $sl(K)$ are
 exchanged 
if $t$ is negative; once again, in the following, I shall assume that $t>0$. 

By taking the logarithm of the Bethe equations and differentiating we
 get a system
of integral equations. 
Let us introduce notations for densities. I call $\rho_p$ $(p=1,N)$ the density per unit length 
of real centers  of $\mu_0$ strings of length $p$  \cplxs (ie the density of
$\mu^{p-1}$ in \cplxs). 
The density of real centers of $l$
strings of $\mu_p$ roots 
that are  not in one of the complexes \cplxs\  we call
$\sigma_p^{(l)}$.  The density of real centers of $l$ strings of $\lambda_p$ 
solutions we call $\tau_p^{(l)}$. I will usually reserve the labels $p,q$ for the colors  of
roots and $l,m$ for the types of strings solutions. We also use the labels $p,q$ for the 
complexes of strings over strings because they behave in many ways
like new roots colors.  

I define the Fourier transform as
\eqn\fourier{\hat{f}(x)=\int d\nu e^{i\nu x}f(\nu),\ f(\nu)={1\over 2\pi}
\int dx e^{-i\nu x}\hat{f}(x),}
and introduce the following notation
\eqn\defi{a_t(\nu)={i\over 2\pi}{d\over d\nu}\ln\left[e_t(\nu)
\right]={1\over 2\pi}{t\over \nu^2+{t^2\over 4}},}
with
\eqn\foui{\hat{a}_t(x)=e^{-t|x|/2}.}
I also define for $r,s$ integers (all these notations are rather standard) 
\eqn\defii{G_{rs}=a_{r+s-2}+a_{r+s-4}+\ldots a_{r-s},\quad r\geq s;\quad G_{rs}=G_{sr},
\quad r\leq s,}
(with $a_0=0$) and
\eqn\defiii{A_{rs}=(G_{rs}+\delta_{rs})\star (1+a_2).}
where $\star$ denotes convolution. Their Fourier transforms are
\eqn\onemorefour{\hat{G}_{rs}={\sinh (xs/2)\over
\sinh (x/2)}e^{-(r-1)|x|/2}-\delta_{rs},\quad  r\geq s,}
and
\eqn\fouii{\hat{A}_{rs}={2\cosh (x/2)\over\sinh(x/2)}\sinh(xs/2) 
e^{-r|x|/2},\quad r\geq s,}
I also introduce the kernel
\eqn\ker{s(\nu)={\pi/2\over \cosh \pi\nu},\ \hat{s}(x)={1\over 2\cosh x/2},}
and define 
\eqn\defiv{a_{rs}=s\star A_{rs},}
with
\eqn\defv{\hat{a}_{rs}(x)={\sinh(xs/2)\over\sinh(x/2)}e^{-r|x|/2},\quad r\geq s.}

I can now write the continuum version of the Bethe equations. Introducing the 
symbol $G_{t+1,p}$ which is defined by a formula similar to \defii\ even when $t$ is not integer
(which is usually the case)
\eqn\defgg{G_{t+1,p}=a_{t+p-1}+a_{t+p-3}+\ldots a_{t-p+1},}
(so, for instance, $G_{2,p}=a_p$) we have, for the
fermionic root, a set of $N$ equations, one for each string 
\eqn\contbethei{G_{t+1,p}=\rho_p+\tilde{\rho}_p +\sum_{q=1}^N G_{pq}\star\rho_q
+\sum_{l\geq 1}a_l\star\sigma_p^{(l)}-\sum_{l\geq 1}a_{pl}
\star\tau_1^{(l)},\quad p=1,\ldots,N.}
For $p=N$ recall that there is no density $\sigma_N^{(l)}$ so the corresponding  term has 
to be suppressed from the equation. For the  $N-1$ roots  of $sl(N)$ we have an infinity of equations,
one for each type of string
\eqn\contbetheii{a_l\star\rho_p=\tilde{\sigma}_p^{(l)}+\sum_{m\geq 1}A_{lm}\star
\sum_{q=1}^{N-1}C_{pq}\star\sigma_q^{(m)},~~ p=1,\ldots,N-1,~~l\geq 1}
where
\eqn\defv{C_{pq}(\nu)=\delta(\nu)\delta_{pq}-s(\nu)\delta_{p,q-1}-s(\nu)\delta_{p,q+1}.}
Finally for the $K-1$ roots of $sl(K)$ roots we have 
\eqn\contbetheiii{\delta_{p1}\sum_{q=1}^N a_{lq}\star\rho_q=\tilde{\tau}_p^{(l)}
+\sum_{m\geq 1}A_{lm}\star\sum_{q=1}^{K-1}C_{pq}\star\tau_q^{(m)},\quad p=1,\ldots,K-1.}

\newsec{The thermodynamic equations}

We now write the thermodynamic equations. The energy term is 
\eqn\ener{E=\sum_{p=1}^N \int \left(\epsilon G_{t+1,p}+pA\right)\rho_p,}
where for notational simplicity, we have not written the variables that are integrated
over (the real centers of the fermionic strings). The chemical potential breaks the 
"supersymmetry", leaving as a symmetry the bosonic part 
$sl(N)\otimes sl(K)\otimes u(1)$. 

We introduce pseudo energies $\epsilon_p,
\kappa_p^{(l)},\zeta_p^{(l)}$ defined by 
\eqn\pseu{\eqalign{\rho_p/\tilde{\rho}_p&=\exp(-\epsilon_p/T),\ p=1,\ldots,N\cr
\sigma_p^{(l)}/\tilde{\sigma}_p^{(l)}&=\exp(-\kappa_p^{(l)}/T),
\ p=1,\ldots,N-1,\ l=1,\ldots,\infty\cr
\tau_p^{(l)}/\tilde{\tau}_p^{(l)}&=\exp(-\zeta_p^{(l)}/T),\ p=1,
\ldots,K-1,\ l=1,\ldots,\infty .\cr}}
The minimization of  the free energy  leads to the system of thermodynamic Bethe ansatz (TBA) equations:
\eqn\tbai{\eqalign{0=&pA-G_{t+1,p}-\epsilon_p+\sum_{q=1}^NG_{pq}\star T\ln
\left(1+e^{-\epsilon_q/T}\right)-\sum_{l\geq 1}a_l\star T\ln
\left(1+e^{-\kappa_p^{(l)}/T}\right)\cr
&-\sum_{l\geq 1}a_{lp}\star\ln
\left(1+e^{-\zeta_1^{(l)}/T}\right),\cr}}
and 
\eqn\tbaii{0=-T\ln\left(1+e^{\kappa_p^{(l)}/T}\right)+\sum_{m\geq 1}
A_{ml}\star
\sum_{q=1}^{N-1}C_{qp}\star T\ln\left(1+e^{-\kappa_q^{(m)}/T}\right)
+a_l\star T\ln\left(1+e^{-\epsilon_p/T}\right),}
and 
\eqn\tbaiii{\eqalign{0=&-T\ln\left(1+e^{\zeta_p^{(l)}/T}\right)+
\sum_{m\geq 1}A_{ml}\star
\sum_{q=1}^{K-1}C_{qp}\star T\ln\left(1+e^{-\zeta_q^{(m)}/T}\right)\cr
&-\delta_{p1}
\sum_{q=1}^N a_{ql}\star T\ln\left(1+e^{-\epsilon_q/T}\right).\cr}}
The free energy reads then
\eqn\freeen{F=-{T\over 2\pi}\int \sum_{p=1}^N G_{p+1,t} \ln\left(1+e^{-\epsilon_p/T}\right
).}

\newsec{Large temperature entropy}

Before going any further, it is useful to check  the completeness of the solutions by studying the large temperature entropy. The
general case is a bit heavy, so I will simply discuss some particular examples here. Let us start 
with $sl(2/1)$, and  introduce the quantities 
\eqn\newquant{\eqalign{x_l=& e^{\kappa_1^{(l)}/T},~~l=1,,\ldots,\infty\cr
y_1=& e^{\epsilon_1/T}\cr
y_2=& e^{\epsilon_2/T}.\cr}}
In the large temperature limit, these go to constants, which are solution of the system
\eqn\system{\eqalign{x_l=& \left[\left(1+x_{l-1}\right)\left(1+x_{l+1}\right)\right]^{1/2} \left(1+{1\over y_1}\right)^{\delta_{l1}/2}\cr
y_1=& \left(1+{1\over y_2}\right)\prod_l \left(1+{1\over x_l}\right)^{-1}\cr
y_2=& \left(1+{1\over y_2}\right)\left(1+{1\over y_1}\right).\cr}}
The solution of this system is 
\eqn\sol{x_l=\left(l+{3\over 2}\right)^2-1,~~y_1={4\over 5},~~y_2=3.}
Meanwhile, at large temperature, one has, for $t\neq 1$
\eqn\largetf{F\approx -T \left[ 2\ln\left(1+{1\over y_2}\right)+\ln\left(1+{1\over y_1}\right)\right]=-T\ln 4,}
in agreement with the dimension of the typical representations of $sl(2/1)$, $d=4$. If, however, $t=1$, one finds
\eqn\largetfi{F\approx -T \left[ \ln\left(1+{1\over y_2}\right)+\ln\left(1+{1\over y
_1}\right)\right]=-T\ln 3,}
in agreement with the dimension of the fermionic atypical representations of $sl(2/1)$, $d=3$. 

wehave performed the same exercise for $sl(3/1)$. It is a bit more laborious, so wewill only give the final result here.
Introducing $y_i=y_1=e^{\epsilon_i/T}$, wefound $y_1={7\over 9}$, $y_2={32\over 17}$, $y_3=7$. The large temperature
free energy for typical representations is then
\eqn\largetfii{F\approx -T \left[ 3\ln\left(1+{1\over y_3}\right)+2\ln\left(1+{1\over y_2}\right)+
\ln\left(1+{1\over y_1}\right)\right]=-T\ln 8.}
There are now two types of atypical representations. If $t=2$, one has
\eqn\largetfiii{F\approx -T \left[ 2\ln\left(1+{1\over y_3}\right)+2\ln\left(1+{1\over y_2}\right)+
\ln\left(1+{1\over y_1}\right)\right]=-T\ln 7,}
while if $t=1$, 
\eqn\largetfiv{F\approx -T \left[ \ln\left(1+{1\over y_3}\right)+\ln\left(1+{1\over y_2}\right)+
\ln\left(1+{1\over y_1}\right)\right]=-T\ln 4.}
All of these coincide with known results of $sl(3/1)$ representation theory. The general relation between the $sl(N/K)$
TBA and representation theory seems quite interesting, but  I won't comment any more on it here \ref\BMR{I. Bars, B. Morel and 
H. Ruegg, J. Math. Phys. 24 (1983) 2253.}.

\newsec{The ground state and physical excitations}

In this paragraph,  I will restrict to the ``generic case'' $t\geq N$. Some special cases
are studied further below.  I will also concentrate on the case $\epsilon=-1$, and comment briefly 
on the case $\epsilon=1$ - which happens to be quite similar -  at the end

\subsec{Equations as $T\rightarrow 0$}

As $T\rightarrow 0$ we find the system  
\eqn\gsi{0=pA-G_{t+1,p}-\epsilon_p-\sum_{q=1}^NG_{pq}\star\epsilon_q^-+
\sum_{l\geq 1}a_l\star\kappa_p^{(l)-}+\sum_{l\geq 1}a_{lp}
\star\zeta_1^{(l)-},}
and
\eqn\gsii{0=-\kappa_p^{(l)+}-\sum_{m\geq 1}A_{ml}
\star\sum_{q=1}^{N-1}
C_{qp}\star
\kappa_q^{(m)-}-a_l\star\epsilon_p^-,}
and
\eqn\gsiii{0=-\zeta_p^{(l)+}-\sum_{m\geq 1}A_{lm}\star\sum_{q=1}^{K-1}C_{qp}
\star\zeta_q^{(m)-}
+\delta_{p1}\sum_{q=1}^N a_{ql}\star\epsilon_q^-.}
where I  have introduced as usual  the positive and negative parts of the
pseudoenergies. Whatever the value of $A$ it is easy to see that one has
\eqn\frii{\epsilon_1^-=\ldots=\epsilon_{N-1}^-=0,}
together with
\eqn\gsxii{\kappa_p^{(l)\pm}=0,\ p=1,\ldots,N-1,
\ l=1,\ldots,\infty.}
and
\eqn\gsxiii{\zeta_p^{(l)+}=0,\quad l=1,\ldots,\infty,;\qquad 
\zeta_p^{(l)-}=0,\quad  l\neq N.}
The equation \gsii\ is then satisfied identically. The equation
 \gsiii\ now reads
\eqn\gsv{0=-A_{lN}\star\sum_{q=1}^{K-1}C_{qp}\star\zeta_q^{(N)-}
+\delta_{p1}a_{Nl}\star\epsilon_N^-.}
Using $a_{Nl}=s\star A_{Nl}$ \defiv, eq. \gsv\ can be rewritten
\eqn\gsvi{0=-\sum_{q=1}^{K-1}C_{qp}\star\zeta_q^{(N)-}
+\delta_{p1}s\star\epsilon_N^-.}
that is, using the form of $C_{pq}$,  
\eqn\gsvii{\eqalign{&\zeta_1^{(N)-}-s\star
\zeta_2^{(N)-}=s\star\epsilon_N^-\cr
&\zeta_2^{(N)-}=s\star\left[\zeta_1^{(N)-}
+\zeta_3^{(N)-}-\right]\cr
&\ldots\cr
&\zeta_{k-1}^{(N)-}=s\star\zeta_{K-2}^{(N)-},\cr}}
whose solution is
\eqn\gsviii{\hat{\zeta}_p^{(N)-}={\sinh(K-p)x/2\over\sinh Kx/2}
\hat{\epsilon}_N^-.}
or
\eqn\friii{\hat{\zeta}_p^{(N)-}=s_{K-p,K}\star\epsilon_N^-,}
where I defined
\eqn\friv{\hat{s}_{rs}={\sinh rx/2\over\sinh sx/2}.}
Similar results hold for the densities, that is 
\eqn\denresi{\rho_1=\ldots=\rho_{N-1}=0,}
together with 
\eqn\denresii{\sigma_p^{(l)}=\tilde{\sigma}_p^{(l)}=0,\quad p=1,\ldots,N-1,\quad l=1,\ldots,\infty,}
and 
\eqn\denresiii{\tilde{\tau}_p^{(l)}=0,\quad p=1,\ldots,K-1,\quad l=1,\ldots,\infty,}
and
\eqn\denresiv{\tau_p^{(l)}=0,\quad p=1,\ldots,K-1,\quad l\neq N.}
From \friii\ we get also
\eqn\denresv{\tau^{(N)}_p=s_{K-p,K}\star\rho_N.}
We thus see that, whatever the value of $A$, the 
only non-vanishing particle
densities are $\rho_N$ and $\tau_p^{(N)}$.  To proceed further we have to distinguish the cases  $A=0$ and $A>0$. 

\subsec{The case $A=0$}

I consider first the case $A=0$. In the ground state one has then, in addition to \frii\
\eqn\gsin{\epsilon_1^{+}=\ldots\epsilon_{N-1}^{+}=0,}
so all the hole densities vanish, leaving the system
\eqn\gsiv{0=-G_{t+1,p}-(\delta_{pN}+G_{pN})\star\epsilon_N^-+a_{Np}\star
\zeta_1^{(N)-},}
From \friii\ we replace $\zeta_1^{(N)-}$ by its expression in 
terms of $\epsilon_1^-$  to get, in terms of 
Fourier transforms,
\eqn\gsix{\eqalign{0=-{\sinh px/2\over\sinh x/2}e^{-t|x|/2}-{\sinh px/2\over\sinh x/2}
e^{-(N-1)|x|/2}\hat{\epsilon}_N^-\cr
+{\sinh px/2\over\sinh x/2}e^{-N|x|/2}
{\sinh(K-1)x/2\over\sinh Kx/2}\hat{\epsilon}_N^-.\cr}}
As expected, $p$ disappears and we get
\eqn\gsx{\hat{\epsilon}_N^-=-{\sinh Kx/2\over\sinh x/2}e^{(N-K-t)|x|/2},\quad 
\hat{\zeta}_p^{(N)-}=-{\sinh (K-p)x/2\over\sinh x/2}
e^{(N-K-t)|x|/2},}
and therefore
\eqn\gsxi{\epsilon_N^-=-a_{t-N+K,K},\quad \zeta_p^{(N)-}=-a_{t-N+K,K-p},}
all other pseudoenergies being zero. In the ground state we therefore have as well, by comparing the Bethe 
equations and the limit $T\rightarrow 0$ of the
thermodynamic equations, 
\eqn\gsxii{\eqalign{\hat{\rho}_N=&{\sinh Kx/2\over\sinh x/2}e^{(N-K-t)|x|/2},\quad
\rho_N=a_{t+K-N,K}\cr
\hat{\tau}_p^{(N)}=&{\sinh (K-p)x/2\over\sinh x/2}
e^{(N-K-t)|x|/2},\quad \tau_p^{(N)}=a_{t+K-N,K-p}\cr},}
all other densities being zero. The ground state is thus filled 
with complexes of $N$ strings over strings 
\cplxs\ and $N$ strings for 
all the $sl(K)$ roots $\lambda_p$.

Excitations  are made of holes in the 
distributions \gsxii, with excitation energies exactly equal to 
 $-\epsilon_N^-$ and $-\zeta_p^{(N)-}$. By a standard argument the momentum is 
given by $p=2\pi\int ({\rm density})$. Taking inverse fourier
 transform we see that these excitation
energies are expressed as  sums of terms of the form $a_t(\nu)$ while momenta
are sum of terms  of the type $i\ln e_t(\nu)$. At large values of the
bare rapidity $\nu$ where the gap vanishes we therefore have
 $e\propto{1\over\nu^2}$ and
$p\propto{1\over\nu}$ ie $e\propto p^2$. The excitations are 
therefore {\sl non relativistic}. 
Such a dispersion relation is characteristic of quantum ferromagnets. However I have
not chosen the "wrong sign" of the hamiltonian: identical features are observed for $\epsilon=1$ (see
below). There does not seem to be a very physical reasons why the excitations 
are not relativistic. Technically, what happens is that
 the fermionic Bethe roots having no self coupling, the dispersion relation of the 
associated dressed excitations is almost the same as the one of the bare excitations. This can be seen 
especially clearly in the case of $sl(1/1)$ (more generally, $N=K$), where the Bethe 
equation reduces to $e_t(\mu_0)^L=1$, and the energy of excitations is $\epsilon_N^-=-a_t$. The hamiltonian
is the one of a XX chain with a magnetic field, $H=\sum_j \left(\sigma_j^+\sigma_j^-+\sigma_j^-\sigma_j^+\right)
-2\sum_j \sigma_j^z$. After fermionization and Fourier transform, it becomes $H=\sum_k 2\cos k a_k^\dagger a_k
-2F$, $F$ the number of fermions. In that language, the ground state is 
obtained by filling up {\sl all} modes $-\pi\leq k\leq \pi$, and the gapless excitations 
occur near $k=0$, where the energy goes like $\epsilon\propto k^2$.

I thus conclude that, if the supersymmetry $sl(N/K)$ is not broken, integrable 
lattice models 
based on fermionic representations do not have a relativistic limit (we will see later
that this is true for other representations as well). This result is 
of course disappointing, and in sharp contrast with the situation for 
ordinary Lie algebras. To get some non trivial results, we do in fact need
to break the supersymmetry, as  I now demonstrate. 

\subsec{The case $A>0$}

Suppose now $A>0$. The first difference with the case $A=0$ is that $\epsilon$ 
has also a non vanishing positive part obeying 
\eqn\gsav{G_{t+1,N}-NA=-\epsilon_N^+-(\delta+G)\star\epsilon_N^-,}
with the kernel
\eqn\gsavi{\hat{G}={\sinh (N-K)x/2\over\sinh Kx/2}e^{-N|x|/2}.}
In particular when $N=K$ we have simply (much like in the $N=K=1$ case)
\eqn\gsavii{G_{t+1,N}-NA=-\epsilon_N^+-\epsilon_N^-,}
The function $\epsilon_N$ is now negative on a finite interval $[-Q,Q]$.
 We have
\eqn\epsili{\epsilon_N^-(\lambda)+\int_{-Q}^Q G(\lambda-\mu)\epsilon_N^-(\mu)=
-G_{t+1,N}+NA,\ \lambda\in [-Q,Q],}
and
\eqn\espilii{\epsilon_N^++\int_{|\mu|\geq Q}H(\lambda-\mu)\epsilon_N^+(\mu)d\mu
=-a_{t-N+K,K}+KA,\ |\lambda|\geq Q}
where $1+\hat{H}={1\over 1+\hat{G}}$. 

These equations can be solved perturbatively in the limit $Q>>1$ using standard
Wiener-Hopf techniques. At dominant order in this limit, the cut-off $Q$ is 
related to the magnetic field by
\eqn\whi{A\approx {1\over K}a_{t-N+K,K}(Q)\approx {{\rm cst}\over Q^2},\ Q>>1,}
(this result is exact in the case $N=K$).  
The system is still gapless but the gap now vanishes at finite rapidity.
 For $\mu$ close
to $\pm Q$ one has
\eqn\massl{\epsilon_N(\mu)\approx |a'_{t-N+K,K}(Q)|(|\mu|-Q),\quad |\mu|\approx Q.}

In the presence of a magnetic field there is no simple relation between the 
ground state
pseudoenergies and densities. The latter obey, instead of \epsili,
\eqn\rhrhi{\rho_N(\lambda)+\int_{-Q}^Q G(\lambda-\mu)\rho_N(\mu)d\mu=G_{t+1,N},
\quad \lambda\in[-Q,Q],}
and
\eqn\rhrhii{\tilde{\rho}_N+\int_{|\mu|\geq Q} H(\lambda-\mu)\tilde{\rho}_N
(\mu)d\mu
=a_{t-N+K,K},\ |\lambda|\geq Q.}
In particular these densities are discontinuous at the cutoff $Q$. For large
$Q$ we have approximately $\rho_N(Q)\approx\tilde{\rho}_N(Q)\approx
 a_{t-N+K,K}(Q)$. On the other hand
the relation between momenta and densities still holds so we get for the 
momentum of excitations \massl\ 
\eqn\momexc{p
_N(\mu)\approx 2\pi a_{t-N+K,K}(Q)(|\mu|-Q),|\mu|\approx Q}
The massless excitations in the $\epsilon_N$ branch therefore  now {\sl are} relativistic, with the sound velocity
\eqn\speed{v_s={1\over 2\pi}{|a'_{t-N+K,K}(Q)|\over a_{t-N+K,K}(Q)}\approx {1\over\pi Q},~~ Q>>1.}

The $\epsilon_N^+$ part induces also non vanishing $\epsilon_p^+$. We have
\eqn\eqeqi{0=pA-G_{t+1,p}-\epsilon_p^+-s_{pN}\star (\delta+G)\star\epsilon_N^-,}
where 
 we used 
$$
\hat{G}_{pN}-\hat{a}_{pN}
{\sinh (K-1)x/2\over\sinh Kx/2}={\sinh px/2\over\sinh Kx/2}e^{(K-N)|x|/2}
$$
Observing that
$$
G_{t+1,p}=
{\sinh px/2\over\sinh Nx/2}G_{t+1,N}
$$
and using \gsav\ we get 
\eqn\simi{\epsilon_p^+=s_{pN}\star \epsilon_N^+,}

Therefore, we have now  new particle like excitations in the system, with energy $\epsilon_p^+$
. Their physical nature is easy to understand. At a given rapidity, when $Q>>1$ only the tails 
at $\pm \infty$ 
 of $s_{pN}$ gives significant contributions because $\epsilon_N^+$ vanishes in $[-Q,Q]$.
 At large argument the behaviour 
of $s_{pN}$ is determined by the pole of its Fourier transform nearest
the real axis, that is $x=2i\pi/N$, and we can approximate 
\eqn\appi{s_{pN}(\nu)\approx {2\over N}\sin\left({p\pi\over N}\right)  
e^{-2\pi|\nu|/N},~~\nu
\rightarrow\pm\infty.}
Expanding $\epsilon_N^+$ close to $Q$, we get therefore
\eqn\dispii{\epsilon_p^+(\mu)\approx {N\over \pi^2}\sin\left({p\pi\over N}\right)
|a_{t-N+K,K}'(Q)|
e^{-2\pi Q/N}\cosh(2\pi\mu/N),\quad |\mu|<<Q,~~ p=1,\ldots,N-1}
The momentum of these excitations is given by $p=2\pi  \int\tilde{\rho}_p$. 
From \eqeqi\ we deduce as well
\eqn\newdeni{\tilde{\rho}_p=s_{pN}\star\tilde{\rho}_N,}
and thus 
\eqn\momi{p={2N\over \pi}\sin\left({p\pi\over N}\right) a_{t-N+K,K}(Q) e^{-2\pi Q/N}\sinh(2\pi\mu/N),\quad |\mu|<<Q.}
Hence the excitations are relativistic once again, 
with a mass
\eqn\mass{m_p={N\over \pi^2}\sin\left({p\pi\over N}\right)
|a_{t-N+K,K}'(Q)|e^{-2\pi Q/N},~~ p=1,\ldots,N-1,}
and the  same sound velocity as before (which was quite obvious from \simi\
and \newdeni.) 

The $\zeta$ excitations are also modified due to the existence of the 
cut-off $Q$. This is easily studied using  \friii\ which still holds. The gap still 
vanishes at rapidities much bigger (in absolute value) than $Q$, but with a different dependence on rapidities.
 In this limit the behaviour
of $\left(\zeta_p^{(N)}\right)^{-}$is determined by the tail of the kernel 
\friv\  
$s_{K-p,K}$. For $\lambda>0\quad (<0)$, only the region close to $Q(-Q)$ contributes, 
so we find
\eqn\dispiii{\left(\zeta_p^{(N)}\right)^{-}\approx {K\over \pi^2}\sin\left({(K-p)\pi\over K}\right)
|a_{t-N+K,K}'(Q)|e^{-2\pi Q/K}e^{\pm 2\pi\mu/K}, \quad|\mu|>> Q.}
The momentum of these excitations is given by $p=2\pi\int \tau_p^{(N)}$.
From \denresv\ we check that these excitations are now left and right moving  relativistic
massless excitations with a  sound velocity that 
is still given by \speed\ and a mass parameter
\eqn\newmomoi{m_p={K\over \pi^2}\sin\left({(K-p)\pi\over K}\right)|a'_{t-N+K,K}(Q)|e^{-2\pi Q/K},~~p=1,\ldots,K-1}
To conclude this section, I would like to notice that
the techniques I used are entirely similar to the ones 
developed by Tsvelik in his study of $sl(2)$ chains with 
a magnetic field \ref\Alold{A. M. Tsvelik, Nucl. Phys. B305 (1988) 675.}

\subsec{The case $\epsilon=1$}

When $\epsilon=1$ and $A=0$, it is easy to see that the solution of the Bethe equations is $\epsilon_p^+=G_{t+1,p}$
for $p=1,\ldots,N-1$, while all other pseudo energies vanish: the ground state is empty,
but there are hole densities for all the fermionic strings. When $A=-B$, $B>0$ is turned on, only $\epsilon_N$ 
acquires a negative part, while the relations $\epsilon_p^+=s_{p,N}\star \epsilon_N^+$ and $\zeta_p^{(N)-}=s_{K-p,K}\star \epsilon_N^-$
still hold. Things are thus very much similar to the case $\epsilon=-1$: the difference is that $\epsilon_N$  is now
 positive (instead of negative)
in a finite interval , and thus it is the $\epsilon_p^+$ excitations that are massless, while the $\zeta_p^{(N)}$ 
excitations are massive. In effect, the roles of $N$ and $K$ are thus exchanged.

\newsec{S matrices in the scaling limit}

We now discuss the way the various excitations interact in the case $\epsilon=-1$ and $A$ a small positive 
number (similar results
would hold for $\epsilon=1$ and $A=-B$, $B$ a small positive number, up to the exchange of $N$ and $K$). In the following we shall be interested in the
scaling limit of the lattice model. In this limit the three types of excitations we
have identified ($\epsilon_p^+,\epsilon_N^\pm,\zeta_p^{(N)-}$) become decoupled
since they occur respectively for rapidities $\mu$ such that
 $|\mu|<<Q,|\mu|\approx Q$
and $|\mu|>>Q$. We shall generally write equations where this coupling has been
neglected by the symbol $\approx$. Densities evaluated in the ground state 
are denoted by $|_0$.

We now get back
to the equations that involve densities (and are magnetic field independent), and consider first the 
physics in the vicinity of $|\mu|<<Q$. In that region, it turns out that physical densities are hole
densities $\tilde{\tau}$, so our first task is to invert \contbetheiii\ and express instead the densities $\tau$ in terms of hole densities
$\tilde{\tau}$. We find
\eqn\inversed{B_{p1}\star s\star\rho_n=  \tau_p^{(n)}- \sum_{q=1}^{K-1}
 B_{pq}\star \sum_{m\geq 1} C^{nm}\star\tilde{\tau}_q^{(m)},~~p,~q=1,\ldots,K-1,}
where
\eqn\defK{B_{pq}={2\cosh x/2\over \sinh x/2 \sinh Kx/2}\sinh\left[(K-p)x/2\right]\sinh[qx/2],~~p\geq q,}
and $C^{lm}$ is defined exactly like in \defv, but acting on upper indices. Also, when $n\geq N$, there is no density $\rho_n$, and the 
source term disappears from the equation \inversed.
The next step is then to replace $\tau_1^{(l)}$ in \contbethei\ by the expression \inversed\ for $p=1$. We then
use the last equation \contbethei\ for $p=N$ to eliminate $\rho_N$. Replacing in the equations \contbethei\ for 
$p=1,\ldots,N-1$ leads to 
\eqn\calci{\eqalign{G_{t+1,p}-\left(G_{t+1,N}\star H_{pN}\star (1+H_{NN})^{-1}\right)+
H_{pN}\star (1+H_{NN})^{-1}\star \tilde{\rho}_N
=\cr
\rho_p+\tilde{\rho}_p+\sum_{q=1}^{N-1}\left(H_{pq}-H_{pN}\star H_{nq}
\star (1+H_{NN})^{-1}\right)\star \rho_q+\sum_{l\geq 1}a_l\star\sigma_p^{(l)}\cr
+\sum_{l\geq 1}\left(H_{pN}\star a_{Nl}\star (1+H_{NN})^{-1}-a_{pl}\right)\star \sum_{q=1}^{K-1}
B_{1q}\star \sum_{m\geq 1}C^{lm}\star
\tilde{\tau}_q^{(m)},\cr}}
where we introduced the kernels
\eqn\newker{H_{pq}={\sinh qx/2\over \sinh Kx/2} e^{(K-p)|x|/2}-\delta_{pq},~~p\geq q.}
The last term vanishes for $l\geq N$; moreover, in the approximation we are considering, it is not possible
to make holes in the distributions $\tau_q^{(N)}$, which would cost a very large energy. The term
$\tilde{\tau}_q^{(N)}$ thus disappears from the equation \calci, leaving a finite set of $\tilde{\tau}$ densities.
 The constant term on the left hand side of \calci\ vanishes identically too.
After simplifying the expressions slightly, we thus end up with the system
\eqn\tdai{\rho_p+\tilde{\rho}_p\approx s_{pN}\star\left.\tilde{\rho}_{N}\right|_0+\sum_{q=1}^{N-1}
Z_{pq}\star\rho_q-\sum_{q=1}^{K-1} s_{K-q,K}\star \tilde{\tau}_q^{(p)} 
-\sum_{l\geq 1}a_l\star\sigma_p^{(l)},~~ p=1,\ldots,N-1}
where
\eqn\tdaii{Z_{rs}=\delta_{rs}-{\sinh rx/2\sinh (N-s)x/2\over \sinh Kx/2\sinh Nx/2}
e^{K|x|/2},\quad r\leq s,\quad Z_{rs}=Z_{sr}.}
This has to be supplemented by the equations 
\eqn\tdaiii{\tilde{\sigma}_p^{(l)}+\sum_{m\geq 1}A_{lm}\star
\sum_{q=1}^{N-1}C_{pq}\star\sigma_q^{(m)}=a_l\star\rho_p,~~ p=1,\ldots,N-1,~~l\geq 1,}
and
\eqn\tdaiv{\tau_q^{(p)}-\sum_{r=1}^{K-1} B_{qr}\star\sum_{s=1}^{N-1} C^{ps}\star \tilde{\tau}_r^{(s)}\approx s_{K-q,K}\star \rho_p,~~p=1,N-1,~~q=1,K-1.}

We can  identify the source terms $ s_{pn}\star\left.\tilde{\rho}_N\right|_0$ with 
$\dot{p}/2\pi$ thanks to \momi. These equations are then easily shown to coincide 
with the system obtained in the study of an $su(N)$ scattering theory at level $K$. Recall in particular that
there are particles of mass $m_p=m\sin(p\pi/N)$ associated with every fully antisymmetric representation
and carrying a charge which is a weight of these representations. The S matrix
is  discussed  in \ref\KR{C. Ahn, D. Bernard and A. Leclair, Nucl. Phys. B346 (1990) 409;
V. V. Bazhanov and  N.Yu Reshetikhin, Prog. Theor. Phys. 102 (1990) 301.}: it is the tensor product of an $sl(N)$ level $K$
RSOS S matrix and a $sl(N)$ ``vertex'' (soliton) S matrix.  This structure is transparent on \tdai: the $\tilde{\tau}_q^{(p)}$ 
are the densities involved in the diagonalization of the RSOS part of the scattering matrix,
while the $\sigma_p^{(l)}$ are those from the diagonalization of the vertex part.

For  the  excitations at rapidities $|\mu|>>Q$, things are a bit simpler, since the 
densities $\rho$ and $\sigma$ are totally frozen in that limit. The physics 
is thus fully described by the equations
\eqn\ciciii{\delta_{p1}a_{lN}\star\left.\rho_N\right|_0\approx^s\tilde{\tau}_p^{(l)}
+\sum_{m\geq 1}A_{lm}\star\sum_{q=1}^{K-1}C_{pq}\star\tau_q^{(m)}.}
These equations  are similar to the ones one would write for a $sl(K)$ lattice model
with the fully symmetric representation $N\omega_1$ on every site, in the limit appropriate
to study the massless right moving excitations \ref\RS{N. Yu Reshetikhin, H. Saleur, Nucl. Phys. B419 (1994) 507}. The scattering
 theory can
then be easily extracted. This time one has massless 
excitations of mass parameter $m_q=m\sin(q\pi/K)$, and the S matrix is the tensor product 
of an $sl(K)$ level $N$ RSOS S matrix and a $sl(K)$ soliton S matrix. This massless theory
is well known to describe the $SU(K)$ level $N$ WZW theory. The idea of describing a conformal 
field theory by a massless scattering theory has a long history going back to \ref\FT{L. D. Fadeev and L. A. Takhtajan, 
Phys. Let. A85 (1981) 375.}. It has been
a subject of intense interest recently in the context  of quantum impurity problems \ref\FSW{
P. Fendley, Phys. Rev. Lett. 71 (1993) 2485; P. Fendley, H. Saleur,
N. P. Warner, Nucl. Phys. B430 (1994) 597} and the quantum KdV equation\ref\BLZ{V. V. Bazhanov, S. L. Lukyanov
 and A. B.  Zamolodchikov, Commun.Math.Phys. 200 (1999) 297-324, and references therein.}.

Finally the $\epsilon_N^{\pm}$ excitations are completely free in this limit, describing a massless $U(1)$ degree
of freedom.

The different pieces of scattering theory found in this section 
coincide with the known results \ref\A{A. Polyakov and P. B. Wiegmann, Phys. Lett. 131B (1983) 121;
P. B. wiegmann, Phys. Lett. 141B (1984) 217.I. Affleck, Nucl. Phys. B265 (1986) 448, and references therein}
for the  $N$ colors, $K$ flavors  chiral Gross Neveu model. We thus 
 conclude that, in the limit of large $Q$ (that is, infinitesimally small symmetry breaking field),
the continuum limit of the $sl(N/K)$ quantum spin chain with  generic fermionic representations obeying $t>N-1$
coincides with the chiral Gross Neveu model 
\eqn\GN{{\cal L}=i\bar{\psi}^{jf}\partial\hskip-.2cm/\psi_{jf}+g \psi_L^{\dagger jf}\psi_{Rjg}\psi_R^{\dagger kg}\psi_{Lkf},}
where $j$ the color index runs from $1$ to $N$ and $f$ the flavor index from $1$ to $K$. In terms of currents,
the interaction reads $J_LJ_R+J_L^aJ_R^a$, where $J$ is the $U(1)$, chirality carrying current,
$J^a$ are the $sl(N)$ currents, with $J_R^a=\psi^{\dagger jf}_R\left(T^a\right)_k^j\psi_{Rkf}$. The parameter $g$ in \GN\ varies 
as $g\propto 1/Q$, and the true scaling limit is obtained when $q\to\infty$, that is $g\to 0$. In that limit, using the equation determining $\epsilon_\pm$, and well
known considerations on dressed charges \ref\Kbook{V. Korepin, N. M. Bogoliubov, and A. G. Izergin, ``Quantum inverse scattering method
and correlation functions'', Cambridge (1993).}, I found that 
the $U(1)$ degree of freedom
has a radius $R={K\over \sqrt{4\pi}}$. Except when $N=K$, this is not 
the radius that is expected for the action \GN, and  the latter requires an
additional $J_LJ_R$ coupling to be correct. As is well known, 
such a coupling does not change any of the physical or integrability properties
in a significant way.

\newsec{Some particular cases}

We concentrated so far on the case of fermionic representations with $t>N-1$ (which, in particular, are
typical).
When $t\leq N-1$, things can get quite  complicated, due to the different possible structures
of the source term $G_{t+1,p}$ in the Bethe equations. 

The simplest situation occurs when $\epsilon=1$. In that case, it is easy to see that when $A=0$, 
the ground state is always given by $\epsilon_p^+=G_{t+1,p}$ (all others being zero), irrespective of the 
value of $t$ (and thus excitations are not relativistic). When the field $A=-B$ is turned on,  $\epsilon_N$  is the only fermionic pesudo energy to 
acquire a negative part, which also gives rise to negative parts $\zeta_p^{(N)-}=s_{K-p,K}\star \epsilon_N^-$. The situation
is thus very similar to the case $t\geq N-1$: the $\epsilon_p^+$ excitations are massless and described 
by an $SU(N)$ level $k$ Wess Zumino model, while the $\zeta_p^{(N)-}$ are massive, and described 
by an $SU(K)$ level $N$ massive theory; the $\epsilon_N$ excitations still  correspond to a simple $U(1)$ theory.
The only difference with the case $t>N-1$ is that the Fermi velocities of these excitations 
are not in general equal anymore. To see this, let us restrict to the case $K=1$ for simplicity. The equations for the
fermionic roots are then
\eqn\partic{\eqalign{0=&-pB+G_{t+1,p}-\epsilon_p^+-G_{pN}\epsilon_N^-,~~p\leq N-1\cr
0=&-NB+G_{t+1,N}-\epsilon_N^+-(1+G_{NN})\epsilon_N^-\cr}}
The pseudo energy $\epsilon_N$ acquires a negative part at large rapidities, while one has
\eqn\junk{\epsilon_p^+={G_{pN}\over 1+G_{NN}} \epsilon_N^++ G_{t+1,p}-{G_{pN}G_{t+1,N}\over 1+G_{NN}}.}
In contrast
with the case $t>N-1$ where the second term vanishes, the massless relativistic region now corresponds to rapidities much larger than $Q$ 
(the Fermi rapidity for $\epsilon_N$), where the first term of \junk\ is negligible: the behaviour of $\epsilon_p$
is thus fully determined by the second term: it is independent of $\epsilon_N$ and $Q$, and thus is bound to have a different
Fermi velocity than the massless $U(1)$ excitations. More careful study of this second term
shows that it reproduces a set of massless excitations with mass parameters proportional to $\sin p\pi/N$, for any value of $t$. The continuum limit of this model is thus the tensor product of a level 1 $SU(N)$ WZW model and a $U(1)$ boson,
each with its own sound velocity.  Notice that it is not necessary to take the $Q\to\infty$ limit here  (since there are no
excitations at small rapidity to decouple)
and as $Q$ varies, the radius of compactification of the boson changes. 
When $Q\to 0$ (low density limit in the t-J language to be discussed below),
it goes  to the point $R=\sqrt{N\over 4\pi}$ of the $c=1$. The whole 
theory thus reproduces the well known system of free fermions with 
$U(1)\times SU(N)$ symmetry  (see
eg \ref\Tbook{A. Tsvelik, ``Quantum field theory in condensed physics'', Cambridge (1995).}). When instead $Q\to\infty$ (corresponding to 
the limit of half filling in the t-J model), the radius goes to $R=\sqrt{1\over 4\pi}$, the same value obtained when bosonizing a single free fermion (which corresponds to 
the case $N=1$). This value arises also when considering the $U\to\infty$ limit
of the $SU(N)$ Hubbard model \ref\Kawa{N. Kawakami, Phys. Rev. B47 (1993) 2928.}.

Indeed, the case $t=1$, $K=1$ is nothing but the so called $SU(N)$ t-J model \LS. It is interesting to notice here that the 
Bethe equations we started with coincide with those of Schlottmann, and therefore to a point of view where there 
is an atypical fermionic representation on every site. As is well known, other Bethe ans\"atze can be written
for this model, as was done first by Sutherland \ref\Su{B. Sutherland, Phys. Rev. B12 (1975) 3795.}: 
they correspond to a point of view where there is a fundamental representation on every site \ref\EK{F. Essler
and V. Korepin, Phys. Rev. B46 (1992) 4197.}. 
 The continuum limit of the t-J model has been worked out in \ref\KY{P. A. Bares and G. Blatter,
 Phys. Rev. Lett. 64 (1990) 2567; N. Kawakami and S. K. Yang, Phys. Rev. Lett. 65 (1990) 2309;
N. Kawamaki, Phys. Rev. B47 (1993) 2928.}: the results are identical to what we found here, although
they are not formulated in terms of massless scattering, but rather using 
conformal dimensions and dressed charge matrices. 

The case $\epsilon=-1$ is considerably more involved, and it is not clear where 
the ground state lies as $t$ varies \RM . An exception  is 
\ $t=N-1$, where, when $A=0$, the ground state is obtained by filling up the sea of length $N-1$ 
fermionic roots, $\epsilon_{N-1}^-=-{\sinh Kx/2\over \sinh x/2}e^{-Kx/2}$, and excitations 
are, as usual, not relativistic. When a field is added,
$\epsilon_{N-1}$ acquires a positive part, together with the other $\epsilon$'s 
which obey $\epsilon_p^+={\sinh px/2\over \sinh (N-1)x/2}\epsilon_{N-1}^+$. The relation
$\zeta_p^{(N-1)-}=s_{K-p,K}\star \epsilon_{N-1}^-$ holds, too. It follows that both $\zeta^{(N-1)-}_p$ and $\epsilon_p^+$  
give rise to relativistic massless (resp. massive) excitations. In addition however, one has 
$\kappa_{N-1}^{(l)+}=-a_l\star \epsilon_{N-1}^-$, giving rise to non relativistic excitations: from the field 
theory point of view, this is thus a not very interesting situation. I  suspect similar 
conclusions hold for other values of $t$. 

\newsec{Other representations}

In the $K=1$ case, the choice  $t=1$ corresponds, in fact, to putting a fundamental representation 
on every site. It is interesting to study more generally
 the case where the chain is built up using  representations
with Dynkin parameters  

\bigskip
\noindent
\vskip.2cm\hskip-3cm\centerline{\hbox{
\rlap{\raise28pt\hbox{\hskip.38cm
$\bigcirc$------$\bigcirc$- - - $\bigotimes$- - - $\bigcirc$------$\bigcirc$}}
\rlap{\raise39pt\hbox{$\hskip .3cm j\hskip.95cm 0\hskip.9cm 0~~\hskip1.75cm
0$}}}
}
\smallskip

\noindent 

The equations then look as in \bethe, except for the source terms: the left hand side for the $\mu_0$ root
is now equal to one, while the left hand side for the $\mu_{N-1}$ root is $e_j^L$, where $j$ is the (integer) Dynkin parameter, $j=a_1$. 
The $su(N)$ t-J model corresponds to $K=1$, $j=1$ already discussed above. In the presence of a chemical potential, the energy reads
\eqn\otherenr{E=-\sum_{\mu_{N-1}}{j\over \mu_{N-1}^2+{j^2\over 4}}+A\sum_{\mu_{N-1}} 1.}

The solutions to the Bethe equations are simpler than in the fermionic case: there are the usual strings for every bosonic root,
while the $\mu_0$ for the fermionic root are all real. With the same notations as before (setting $\rho_1\equiv \rho$)
we now have, going to the continuum version
\eqn\atybethe{G_{j+1,l}\delta_{p,N-1}=\tilde{\sigma}_p^{(l)}+\sum_{m\geq 1}A_{lm}\star\sum_{q=1}^{N-1}C_{pq}\star\sigma_q^{(m)}-\delta_{p1}
a_l\star\rho,}
together with 
\eqn\atybethei{0=\rho+\tilde{\rho}-\sum_{l\geq 1} a_l\star\sigma_1^{(l)}+\sum_{l\geq 1} a_l\star \tau_1^{(l)},~~p=1,\ldots,N-1,}
and 
\eqn\atybetheii{\delta_{p1}a_l\star\rho=\tilde{\tau}_p^{(l)}+\sum_{m\geq 1} a_{lm}\star\sum_{q=1}^{K-1} C_{pq}\star\tau_q^{(m)},~~p=1,\ldots,K-1.}

To discuss what is going on, let us consider as an example the case $N=K=2$. The TBA equations 
 at $T=0$ are, setting $\kappa_1^{(l)}\equiv \kappa^{(l)}$, $\zeta_1^{(l)}=\zeta^{(l)}$, 
\eqn\atytba{\eqalign{G_{j+1,l}=&-\kappa^{(l)+}-\sum A_{ml}\star\kappa^{(m)-} + a_l\star\epsilon^-\cr
0=&A-\epsilon+\sum a_l \star\kappa^{(l)-}-\sum a_l \star\zeta^{(l)-}\cr
0=&-\zeta^{(l)+}-\sum A_{lm}\star \zeta^{(m)-}+a_l\star\epsilon^-.\cr}}
This system can be transformed into
\eqn\atytabi{\eqalign{\kappa^{(l)}=&s\star\left(\kappa^{(l-1)+}+\kappa^{(l+1)+}\right)+ \delta_{l1} s\star\epsilon^--\delta_{lj} s\cr
\epsilon=&A+\sum a_l\star\kappa^{(l)-}-\sum a_l\star\zeta^{(l)-}\cr
\zeta^{(l)}=&s\star\left(\zeta^{(l-1)+}+\zeta^{(l+1)+}\right)+ \delta_{l1} s\star\epsilon^-.\cr}}
In the ground state, all positive parts of the pseudo energies vanish. It follows that $\kappa^{(1)-}=s\star\epsilon^-$ and $\kappa^{(j)-}=-s$,
$\zeta^{(1)-}=s\star\epsilon^-$.  These can be put back in the equation for $\epsilon$, which reads in that case (it is particularly 
simple because $N=K$ here)
\eqn\atytbaii{\epsilon=A- s\star a_j.}
When $A$ vanishes, the $\epsilon$  excitations are non relativistic. When $A$ is positive,
and provided it is not too big,  $\epsilon$ has a positive and a negative part; the corresponding $u(1)$ excitations are
then  massless; calling $Q$ the 
Fermi rapidity for the $\epsilon$ excitations, their sound velocity is $v={1\over 2\pi} {|\epsilon'(Q)|\over
\rho(Q)}$.  From \atytabi, it also follows that $\kappa^{(j)-}=-s$,
$\kappa^{(1)-}=s\star\epsilon^-$ and $\zeta^{(1)-}=s\star\epsilon^-$. The $\kappa$ excitations  are thus 
also massless and relativistic 
at rapidities much larger than $Q$: notice however that they have 
different Fermi velocities: the one for $\kappa^{(j)-}$ is independent of $Q$ (it turns out to be $v_j=\pi$ in our conventions)
, while the one for $\kappa^{(1)-}$ does
depend on it, and reads $v_1={1\over 2\pi}{|\kappa^{(1)-'}(\mu)|\over \sigma^{(1)}(\mu)},\mu\to\infty$. 
As for the $\zeta$ excitations, they are also massless, with a similar sound velocity. 

Much like in the previous section, and in contrast with the case of generic
fermionic representations (with $t\geq N-1$),  relativistic invariance 
does not require $Q$ to be large, and thus 
is obtained for an entire range of values of $A$. Here, this means that the associated scattering 
theory has a continuous parameter $A$, whose meaning we now partly elucidate.
To do so, we observe that the equations  for the densities  are  identical to a decoupled system 
for a pair of $sl(2)$  spin chains, the first $sl(2)$ chain having a source term on the first  and $j^{th}$ node,
the second $sl(2)$ chain on the first node. The second system has thus the $su(2)$ level 1 
WZW model as continuum limit. As for the first, it is similar  to a general class of lattice models
with mixtures of several spins. These models were discussed in details in 
\ref\ADJ{N. Andrei, M. Douglas and A. Jerez, cond-mat/9502082.} and 
 \ref\SS{H. Saleur, P. Simonetti, Nucl. Phys. B 535 (1998) 596}. In \SS, the sound velocities for all excitations were the same, 
and the  
continuum limit was  described as the tensor product of an  $SU(2)$ level $j$ WZW model (with $c={3j\over j+2}$)
and of an $SU(2)$ minimal coset model with $c=1-{6\over (j+1)(j+2)}$. In the limit of small $Q$, 
this result essentially still holds, but now the two types of excitations  each have different Fermi velocities, 
$v_j$ and $v_1$ respectively. The scattering theory is as in \SS, and the field theory can be related with 
a 2 colors $j$ flavors Gross Neveu model with flavor anisotropy. 
 Away from that limit, things are more complicated: the contribution to the free energy from
the $\kappa$ degrees of freedom can be written as $f_{\kappa}\approx -{\pi T^2\over 6}\left({c_j\over v_j}+{c_1\over v_1}\right)$, 
and although the sum $c_j+c_1$ stays the same (it is controlled by the overall shape of the 
TBA diagram), the individual  values of these two parameters 
evolve with $A$. In the limit where $Q\to\infty$, one finds that $c_1=1$ 
while $c_j={3(j-1)\over j+1}$. One should not however think that this model
always decomposes into the sum of two independent CFTs with different sound velocities,
and central charges $c_j$ and $c_1$, except for very small and very large $Q$.
To get an idea
of what happens, I will restrict to  the case $j=2$, which was also partly treated in \FPT.  Consider 
therefore a theory made initially of an Ising model and a level 2 WZW model, which 
is the right description of the system at small $A$. This theory  can be written in 
terms of 4 species of Majorana fermions $\chi_i$, $i=0,1,2,3$. Suppose now one adds,
as suggested in \FPT, a four fermion coupling in each chiral sector. The hamiltonian 
say for the right movers reads then
$$
{\cal H}=-i a\sum_{i=1}^3 \chi_i\partial_x\chi_i-ib\chi_0\partial_x\chi_0+c\chi_0\chi_1\chi_2\chi_3.
$$
To handle this model, the best is to bosonize, representing the fermions as $\chi_0\propto\cos\phi_1,\chi_1\propto\sin\phi_1$ \ref\KK{E. Kiritsis, Phys. Lett. B198 (2987) 379.} (and similarly for $\chi_2,\chi_3$ in terms of another boson $\phi_2$).  The hamiltonian reads then, schematically,
$$
{\cal H}=A\left[\left(\partial\phi_1\right)^2+2\left(\partial\phi_2\right)^2+i\alpha_0\partial^2\phi_1\right]
+B\left[\left(\partial\phi_1\right)^2-i\alpha_0\partial^2\phi_1\right]
+C\partial\phi_1\partial\phi_2
$$
where $\partial\equiv\partial_x$. For $C=0$, and a choice of $\alpha_0$ corresponding 
to the bosonized Ising stress energy tensor, the $A$ term is a $c=3/2$ theory, the $B$ term 
a $c=1/2$ theory. Moreover, these two theories are independent (the short distance expansion
of their two stress energy tensors is regular). For $\alpha_0=0$, and $A=B$, the theory 
decomposes in two independent bosons $\phi_1\pm \phi_2$ with different 
sound velocities, obtained by diagonalizing the quadratic form. In general however, 
the theory defined by ${\cal H}$ is not the sum of two independent conformal
field theories. The massless excitations 
identified previously have to be thought of 
as a way to define the excitations of the whole theory, 
which is therefore  not conformal invariant, since it does not have a well defined sound velocity. 
It would definitely be of interest to study the finite size spectrum of the lattice
model to push this investigation further.  

The foregoing results generalize easily to arbitrary $N,K$. The key point is that 
the $\epsilon$ excitations give rise to a massless $U(1)$ (charge) degree of freedom, and that this 
excitation in turn feeds source terms on the $l=1$ strings for both the (color) $sl(N)$ and  (flavor) $sl(K)$ 
excitations. As a result, one always gets the  $SU(K)$ level one WZW model in the flavor sector, and a more complex 
theory  in the color sector, that reduces 
to a mixture of the $SU(N)$ level
$j$ WZW model and an $SU(N)$ coset model in limit of small $A$.

\newsec{Impurities}

By a very general construction, it is possible to build an integrable impurity model by inserting a different 
representation, or a representation with a different spectral parameter, in the general quantum inverse scattering
framework. This is the same trick that has often been used to study models with mixtures of 
representations, as well as models with spectral parameter heterogeneities \ref\misc{L. D. Faddeev and N. Yu Reshetikhin, Ann. Phys. 167 (1986) 227; A. G. Izergin and v. E. Korepin, Lett. Math. Phys. 5 (1981) 199;
C. Destri and H. de Vega, J. Phys. A22 (1989) 1329.},\RS.  In the context of impurities, the method was probably 
first used in \ref\AJ{N. Andrei and H. Johannesson, Phys. Lett. A100 (1984) 108}. It was applied for instance to the t-J model
with a four dimensional impurity (the $sl(2/1)$ case) in \ref\BEF{G. Bed\"urftig, F. H. Essler, and H. Frahm,
Nucl. Phys. B489 (1997) 697.}.

Note that I am only discussing impurities in a periodic chain here; this is not the same (although
results in the continuum limit are quite related) than having a chain with eg open boundaries,
nor boundary impurities. For some recent results in that direction, see eg \ref\ZG{H. Q. Zhou and 
M. D. Gould, cond-mat/9809055; H. Fan, B. Y. Hou and K. J. Shi, cond-mat/9803215.}.

\subsec{Fermionic impurities in a generic fermionic chain}

Let us first suppose that we insert in a chain based on fermionic representations with $t>N-1$,
 a fermionic representation with a different value of the Dynkin parameter, and a shifted 
spectral parameter. The Bethe equations now look like \bethe, except that for the $\mu_0$ root, the left hand side
contains an additional term $e_{t'}(\mu_0-\Upsilon)$. The energy takes the same form as before, and all the impurity 
term changes is the equations for the densities: equations \contbetheii\ and \contbetheiii\ are unchanged,
while \contbethei\ contains an additional ${1\over L} G_{t'+1,p}(\mu_0-\Upsilon)$, $L$ the size of the system.

Since the equations for the densities appear in the thermodynamic Bethe equations only through their variations, the
equations determining the ground state are unchanged: \gsi,\gsii\ and \gsiii\ still hold, together with the 
analysis of the previous sections.  Proceeding further to analyze the scattering of the 
excitations, equations \tdai,\tdaiii\ and \ciciii\ still hold, too. The only role of the impurity 
is that $\rho_N$ and $\tilde{\rho}_N$ in the left hand sides of \tdai\ and \ciciii\ have now a $1/L$ part,
which follows from the solution of the equation generalizing \rhrhi:
\eqn\onlynew{\rho_N(\lambda)+\int_{-Q}^Q K(\lambda-\mu)\rho_N(\mu)d\mu=G_{t+1,N}(\lambda)+{1\over L} G_{t'+1,N}(\lambda-\Upsilon).}
Note that the Fermi cut-off $Q$ is not changed, since it follows from the condition $\epsilon_N^\pm(Q)=0$,
and the latter is a TBA equation, independent of any impurity terms. The Fermi velocity 
has a $1/L$ correction that we neglect in the limit $L$ large. The only effect of the impurity is thus 
to modify $\tilde{\rho}_N$ and $\rho_N$ in the equations \tdai\ and \ciciii\ by a term of order $1/L$. Consider
for instance equation \tdai. Since $\tilde{\rho}_N$ still vanishes for rapidities smaller or equal to $Q$, the impurity term
does not introduce any non trivial phase shift for the densities $\rho_p$;  its only effect at leading order in the scaling limit $Q\to\infty$  is to renormalize the mass of the excitations (keeping 
their ratios constant) by a $1/L$ term which becomes negligible in the limit $L$ 
large.  The impurity   does not introduce any flow in the renormalization
group sense, and  roughly corresponds to changing the length of the system
by a finite amount (I'll refer to this, not quite correctly,  as being ``irrelevant'') .  The same conclusion holds for the $u(1)$ sector.

In contrast, suppose now that $t'<N-1$. In that case, as has been noticed before,  
the combination 
$$
G_{t'+1,p}-{H_{pN}G_{t'+1,N}\over 1+H_{NN}}=G_{t'+1,p}-{G_{pN}G_{t'+1,N}\over 1+G_{NN}}
$$ 
does not vanish. Calling this combination $I_{t',p}$,
it follows that for large $L$ the right hand side of equation \tdai\ contains now two source terms, $\left. s_{pN}\star \tilde{\rho}_N\right|_0$
and ${1\over L} I_{t',p}(\mu-\Upsilon)$. 

In the case $t'=1$, the impurity term  is ${1\over L}s_{p,N}(\mu-\Upsilon)$: the equations exactly 
coincide with those of  the exactly screened $su(N)$ Kondo model with $K$ channels \ref\TW{A. M. Tsvelik and P. Wiegmann, Adv. Phys. 32 (1983) 453},\ref\Arev{N. Andrei, K. Furuya and J. Lowenstein, Rev. Mod. Phys. 55 (1983) 331.}, although the bulk
now appears massive. If however one concentrates on the massless limit of the bulk degrees of freedom
by letting the rapidity $\mu\to\infty$, while at the same time also sending 
the impurity rapidity $\Upsilon\to\infty$, the Kondo equations are then 
exactly recovered, $2\pi\Upsilon/N$ being related in a simple way to  the Kondo rapidity.

In fact, it is easy to see that  our   Bethe equations coincide with  those of the degenerate Anderson
model when $K=1$ - this has already been observed by Schlottmann in some cases \ref\Simp{P. Schlottmann, 
Z. Phys. B49 (1982) 109.}. Indeed, the equations for the degenerate Anderson model as written
say in \TW\ are the same as the ones we are considering, 
with $t'=1$, $\Upsilon={\epsilon_d\over 2\Gamma}$, $\mu_0={k_j\over 2\Gamma}$, {\sl except} that the source term
for the Anderson model is $\exp(i k_jL)\equiv \exp(2i\Gamma\mu_0 L)$ instead of our $[e_t(\mu_0)]^L$. To
match these too, it  is enough  to send $t\to\infty$ with $t\propto 1/\Gamma$, and rescale the length
of the system appropriately. Since the physical properties are independent of $t$ for $t>N-1$, we should indeed
obtain the same results as for the degenerate Anderson model when $t'=1$. 

Introducing the physical rapidity $\theta={2\pi \mu\over N}$, such that the dispersion relation of 
massless excitations is $p=e\propto e^\theta$,  we find that
\eqn\Konscat{s_{N-p,N}(\theta)={1\over i}{d\over d\theta} {\sin\left({\theta\over 2i}-{\pi p\over 2N}\right)
\over \sin\left({\theta\over 2i}+{\pi p\over 2N}\right)}\equiv {1\over i}{d\over d\theta} (p).}

In the case of higher values of $t'$, we find more complex reflexion matrices. For $t'=2$ for instance, one has $(p-1)(p+1)$,
etc. The meaning of these scattering matrices will be discussed further in \ref\FS{P. Fendley, H. Saleur,`` Integrable
boundary theories with 
 $SU(n)$ symmetry'', in preparation.}. 

To conclude, we see that, while the continuum limit without impurity was 
a Gross Neveu model with $N$ colors and $K$ flavors, either the fermionic impurity is irrelevant if $t'>N-1$, or it affects the $su(N)$ sector 
in the same way as in a pure $su(N)$ theory (as for the $su(K)$ and $u(1)$ sector, they are still unaffected). 

Non fermionic impurities in the fermionic chain could also be considered, but their effect is quite 
straightforward: they affect the $SU(N)$ or $SU(K)$ sectors of the Gross Neveu model
as in pure $SU(N)$ or $SU(K)$ theories,
giving rise to various $N$ or $K$ channel Kondo models (a solid state physics model 
corresponding to this situation was proposed in \ref\K{I.N. Karnaukhov, ``Integrable 
model of orbital degenerate electrons interacting with impurity'', preprint March 15, 1995.}). 

\subsec{Fermionic impurities in a non fermionic chain}

Another example of interest is a fermionic impurity in a non fermionic chain. Consider for instance 
again the case $N=K=2$ with a representation having Dynkin parameters $a_1=j$ in the bulk. The only difference 
with the analysis of section 8 is that a $1/L$ term appears in the right hand side of equation \atybethei: the situation
is thus very  similar to the case of a fermionic representation in a fermionic chain: the impurity renormalizes the 
source terms for the two $SU(2)$ systems by a term of order $1/L$ is 
rapidity independent: no flow is generated. This conclusion generalizes to any $N,K$, and seems to agree with the results of \BEF.

\newsec{Conclusion}

In conclusion, it does not seem possible to observe any interesting ``supersymmetric'' properties
 in the continuum limit of integrable lattice models based on $sl(N/K)$ superalgebras
\foot{Models based on $osp(N/2M)$ seem more promising,
since, according to \ref\MNR{M. J. Martins, B. Nienhuis, R. Rietman, Phys. Rev. Lett. 81 (1998) 504.},
the ones based on the fundamental representation {\sl are} conformal invariant.}  
. In fact,
 with the whole superalgebra symmetry, the continuum limit of these chains is not even relativistic,
 in sharp contrast with what happens in the case of ordinary algebras, where this continuum limit
 coincides with Wess Zumino models on the corresponding group. Interesting continuum limits can 
 be obtained only when the superalgebra symmetry is broken. The case we have considered in details here
 leaves the $sl(N)\otimes sl(K)\otimes u(1)$ symmetry, and, in the continuum limit, gives various
 instances of color and flavor Gross Neveu models. The most ``symmetric'' case is obtained 
 with fermionic representations with Dynkin parameter $t\geq N-1$: in that case, one truly gets the 
 $N$ colors, $K$ flavors, Gross Neveu model, with the remarkable property that all the excitations 
 have the same sound velocity indeed. Other cases lead to continuum field theories with less symmetry, 
 typically involving mixtures of massive and massless excitations with different sound velocities. We have also 
 considered impurity models, concluding that in that case too, nothing really new is observed, impurities
 either  leading to irrelevant perturbations, or reproducing known Kondo models
 in the $SU(N)$ or $SU(K)$ sectors of the GN model. In particular, the continuous parameter 
that is our disposal when using typical representations does not give rise to 
interesting tunable parameters in the field theory limit; in general, it simply 
affects the sound velocity, or the overall mass scale. 

It must be stressed that models based on the quantum deformations $sl_q(N/K)$ {\sl would} be relativistic, 
even without the introduction of a chemical potential (this is especially easy to 
see in the case of $sl(1/1)$, where turning on the quantum group deformation 
is equivalent to adding up the chemical potential $A$). It is not known what
the continuum limit of these models is in general, nor what happens to them as $q\to 1$; this is actually
a very intriguing question, on which I hope to report soon. 

An issue that is somewhat related is what would happen for models based on, for instance,  alternating
fundamental representation and its conjugate. Unlike in the $sl(N)$ case, the conjugate of the fundamental
does not behave like another fundamental representation. While, for instance, the $sl(3)$ model
with an alternance of $3$ and $\bar{3}$ is integrable, I do not know whether the $sl(2/1)$ model
is, nor what the Bethe equations would look like. This problem deserves more study,
as it seems related with important issues in disordered systems \ref\GLR{I. Gruzberg, A. Ludwig and N. Read, cond-mat/9902063.}. 

\bigskip
\noindent{\bf Acknowledgments}: I am grateful to A. Tsvelik, who kindly
encouraged me 
to publish these results despite their somewhat disappointing nature. I  also would like to 
thank N. Yu Reshetikhin for interesting discussions on super algebras and the Bethe ansatz. This work was supported by the DOE and the NSF (under the NYI program). 
\listrefs

\bye